\documentclass[preprint,superscriptaddress]{revtex4-1}
\usepackage{graphicx}
\usepackage{braket}
\usepackage{amssymb,amsmath}
\usepackage[T1]{fontenc}
\usepackage[english]{babel}
\usepackage{textgreek}

\begin{document}

%Title of paper
\title{Phase cycling of extreme ultraviolet pulse sequences generated in rare gases}
\author{Andreas Wituschek}
\affiliation{Institute of Physics, University of Freiburg, Hermann-Herder-Str. 3, 79104 Freiburg, Germany}
\author{Oleg Kornilov}
\author{Tobias Witting}
\author{Laura Maikowski}
\affiliation{Max-Born-Institut, Max-Born-Str. 2A, 12489 Berlin, Germany}
\author{Frank Stienkemeier}
\affiliation{Institute of Physics, University of Freiburg, Hermann-Herder-Str. 3, 79104 Freiburg, Germany}
\author{Marc J.J. Vrakking}
\affiliation{Max-Born-Institut, Max-Born-Str. 2A, 12489 Berlin, Germany}
\author{Lukas Bruder}
\email{lukas.bruder@physik.uni-freiburg.de}
\affiliation{Institute of Physics, University of Freiburg, Hermann-Herder-Str. 3, 79104 Freiburg, Germany}

\date{November 02, 2020}

\begin{abstract}
The development of schemes for coherent nonlinear time-domain spectroscopy in the extreme-ultraviolet regime (XUV) has so far been impeded by experimental difficulties that arise at these short wavelengths. 
In this work we present a novel experimental approach, which facilitates the timing control and phase cycling of XUV pulse sequences produced by harmonic generation in rare gases. 
The method is demonstrated for the generation and high spectral resolution characterization of narrow-bandwidth harmonics ($\approx$14\,eV) in argon and krypton.
Our technique simultaneously provides high phase stability and a pathway-selective detection scheme for nonlinear signals - both necessary prerequisites for all types of coherent nonlinear spectroscopy.
\end{abstract}

\maketitle

%%%%%%%%%%%%%%%%%%%%%%%%%%  body  %%%%%%%%%%%%%%%%%%%%%%%%%%

\section{Introduction}
Coherent nonlinear spectroscopy in the time domain is a powerful tool to study photoinduced dynamics in complex quantum systems on their natural time scale\,\cite{mukamel_principles_1999}.
An extension to the XUV spectral regime is highly desirable, as it would in principle foster studies with attosecond temporal resolution and site or chemical selectivity\,\cite{mukamel_multidimensional_2013,kraus_ultrafast_2018}.
In coherent nonlinear spectroscopy, sequences of phase-locked ultrashort laser pulses interact with a system, simultaneously exciting many quantum pathways.
Subsequently, observables are measured as a function of the time delay between the pulses, which in combination with Fourier-transform allows to observe spectral signatures. 
%This requires the generation of phase-locked pulse sequences. 
%On the other hand, in nonlinear spectroscopy many quantum pathways are simultaneously excited.
%Hence, to extract the relevant spectroscopic information, highly sensitive pathway-selective detection methods are essential\,\cite{mukamel_principles_1999,hamm_concepts_2011}. %, e.g.~to facilitate a selective detection of a single high harmonic.
Therefore these methods require generation of phase-locked pulse sequences and highly sensitive pathway-selective detection methods\,\cite{mukamel_principles_1999,hamm_concepts_2011}.
While this is readily achieved in the visible regime, the simultaneous experimental realization of both ingredients constitutes a major challenge at XUV wavelengths.

Phase-locking in the visible regime is typically achieved using a passively or actively stabilized interferometric setup, which splits, delays and recombines the pulses.
At XUV wavelengths fundamental mechanical constraints limit the phase stability of most implementations and technical difficulties (for example the lack of transmissive optics and absorption of XUV light in air) impede the experimental realization.
Therefore only few studies have demonstrated phase-locked XUV pulses.
Among those, at free-electron lasers (FELs) phase-locking was achieved using double-pulse seeding with phase-locked pulses\,\cite{gauthier_generation_2016}, by manipulating the trajectory of the relativistic electron bunch used for generation of the XUV radiation\,\cite{prince_coherent_2016}, or by elaborate interferometric setups directly in the XUV\,\cite{usenko_attosecond_2017}.
Ultimately phase-locking of several harmonics of a FEL, has led to generation of attosecond pulses at the \textmu J level\,\cite{maroju_attosecond_2020}.
In the case of tabletop High Harmonic Generation (HHG), phase-locked XUV pulse-sequences were generated using phase-locked double-pulse pumping\,\cite{cavalieri_ramsey-type_2002,jansen_spatially_2016}, direct manipulation in a monolithic interferometric setup\,\cite{okino_direct_2015}, in actively stabilized XUV Michelson-type interferometers with wavefront splitting optics\,\cite{yost_vacuum-ultraviolet_2009}, and by placing two movable gas targets within the Rayleigh-range of the pump laser\,\cite{laban_extreme_2012}. 
A particularly sophisticated implementation uses separately amplified pulses from a frequency comb\,\cite{witte_deep-ultraviolet_2005}.
However, none of the mentioned examples has facilitated pathway-selective detection in parallel.

Pathway-selective detection in the XUV regime has been achieved in four-wave mixing (FWM) geometries.
In FWM multiple pulses with individual timing are directed non-collinearly at the sample \,\cite{hamm_concepts_2011}.
The nonlinear mixing signals are then radiated in specific directions and can be isolated by spatial filtering, yielding sensitive, background-free detection. 
Examples are the FWM beamline at the FERMI FEL\,\cite{foglia_first_2018, bencivenga_four-wave_2015} and experiments exploiting the inherent synchronization between the NIR driving pulse and the resulting XUV pulse in HHG schemes\,\cite{marroux_multidimensional_2018,warrick_multiple_2018}.
However, owing to the complex geometries of the FWM setup, the phase stability is limited and the overall sensitivity is constrained by straylight since FWM is restricted to detection of photons.

In a recent study we have simultaneously implemented XUV phase-locking and pathway-selective detection using double-pulse pumping of a FEL with phase-modulated pulse sequences\,\cite{wituschek_tracking_2020}. 
This enabled tracking of inner-subshell valence-shell electronic coherences spanning over 28\,eV and efficient background suppression.
The phase-modulation technique is a highly sensitive technique to record electronic wave-packet interferences.
It was originally developed in the visible regime\,\cite{tekavec_wave_2006}, but extensions to UV wavelengths\,\cite{widom_solution_2013, bruder_phase-modulated_2017,wituschek_stable_2019} and to an XUV FEL have been demonstrated\,\cite{wituschek_tracking_2020}.
In a nutshell, the relative phase of the excitation pulses is cycled on a shot-to-shot basis, leading to distinct modulation patterns in the (non)linear response of the system, which are monitored using incoherent 'action'-signals like fluorescence\,\cite{tekavec_wave_2006} or photoionization yields\,\cite{bruder_phase-modulated_2015}.
These distinct modulations are readily demodulated using a lock-in amplifier, allowing for isolation of specific excitation pathways (i.e. pathway-selective detection).
Due to the lock-in detection the method is highly sensitive. 
For example, it enabled isolation of dipole interactions in extremely dilute atomic vapors\,\cite{bruder_delocalized_2019}.
Furthermore, a straightforward extension allows for two-dimensional electronic spectroscopy\,\cite{tekavec_fluorescence-detected_2007,bruder_coherent_2018}, which is a powerful technique to follow electronic dynamics in real-time.

In this work, we combine the phase-modulation technique for the first time with harmonic generation in rare gases, yielding high phase stability and pathway selectivity in a single setup. 
Our results imply that the shot-to-shot phase cycling in the pump pulses is coherently transferred to the XUV pulses.
By performing high-resolution linear interferometric cross correlations (CCs) of the XUV pulses, we find that the resulting XUV radiation is extremely narrowband.
Calculations show that this is a consequence of enhanced phase matching on the high energy side of atomic resonances. 
Our method provides high phase stability and the necessary pathway-selective detection scheme to isolate the weak CC signals, which upon Fourier-transform, provides spectral information about the XUV light.
In the present case, the high resolution of the Fourier-transform approach allows us to identify the contribution of closely spaced individual resonances to the overall spectrum.

\section{Methods}
\subsection{Experimental setup}
\begin{figure}
\centering\includegraphics[width=0.8\linewidth]{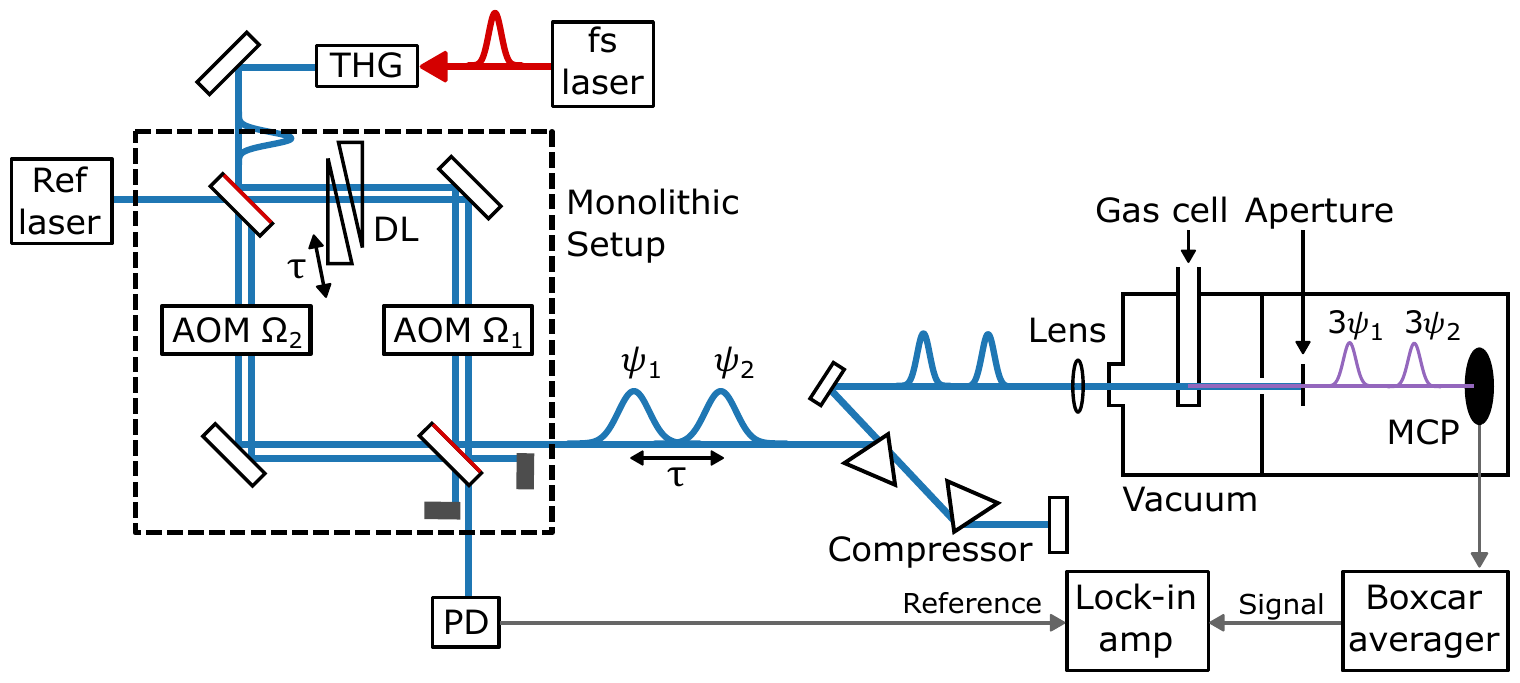}
\caption{Third harmonic generation (THG) of NIR pulses yields femtosecond UV pulses at 266\,nm which are split in an interferometric setup. The beamsplitters have a 50/50 coating which is indicated in red. The delay $\tau$ is controlled with a wedge-based delay line (DL), the phase $\psi_{i} =\Omega_i t$ is controlled using acousto-optical modulators (AOMs) which are driven on distinct phase-locked radio frequencies $\Omega_i$. 
The interferometer is traced with a continuous wave reference laser. Its interference is recorded with a photodiode (PD) and is used as a reference for the lock-in amplifier. The UV pulses are compressed using a CaF\textsubscript{2} prism compressor and focused into the gas cell. Residual UV light is reduced using an aperture. The intensity of the XUV light is monitored using a microchannel plate (MCP), whose output is fed into a boxcar averager and subsequently to the lock-in signal port.}
\label{fig:setup}
\end{figure}

The experimental setup is shown in figure\,\ref{fig:setup}.
The laser system (Amplitude Technologies) provides 24\,fs pulses centered at 790-800\,nm at a repetition rate of 1\,kHz with pulse energies up to 6\,mJ.
For timing control and phase modulation of the intense femtosecond pulses, we employ a specialized monolithic setup that can handle high pulse energies at wavelengths of $\approx 266\,$nm, i.e. the third harmonic of the fundamental wavelength of the laser.
The setup was previously introduced in Ref.\,\cite{wituschek_stable_2019}. 
%Therefore, a set of nonlinear crystals for third harmonic generation (THG) prior to the phase modulation setup was used (Eksma Optics).
The 266\,nm pulses are generated by a commercial third-harmonic generation (THG) kit (Eksma Optics).
The residual near-infrared (NIR) and second harmonic light are removed using dichroic mirrors. 
Wavelength tuning of the resulting ultraviolet (UV) pulses between 264 - 268\,nm is achieved by tuning of the fundamental NIR wavelength and simultaneously adjusting the phase-matching angles of the THG crystals.
%The resulting UV pulses were split, delayed and recombined using a highly stable interferometric platform\,\cite{wituschek_stable_2019}. 
%Note that the beam path in the interferometer is not actively stabilized, in contrast to other double-pulse pumping studies\,\cite{corsi_ultrastable_2015}.
%Unless otherwise stated, the pulses were recombined in a collinear fashion.
In the phase-modulation setup (see figure\,\ref{fig:setup}) the UV beam is split by a 50/50 beamsplitter and the relative phase between both arms is modulated with acousto-optical modulators (AOMs).
The delay $\tau$ between the two pulses is controlled by a pair of fused-silica wedges.
The position of one of the wedges is motorized, which allows to scan the delay from -1\,ps to  11\,ps.
To balance the optical path introduced by the wedges, the beam in the arm without the wedges passes through the glass of each beamsplitter, while the other beam is reflected from each beamsplitter (see figure\,\ref{fig:setup}).
The delay was calibrated in an earlier work using the known transition frequency of an atomic resonance and the known reference laser frequency\,\cite{wituschek_stable_2019}.

Unless otherwise stated, the pulses are recombined in a collinear fashion.
Note that the beam path in the interferometer is not actively stabilized, in contrast to other double-pulse pumping studies\,\cite{corsi_ultrastable_2015}.
%The beams in both arms of the interferometer pass through AOMs, which enables control over the relative phase of the two beams.
The total path-length in the bulk of the optics of the interferometer (AOMs, beamsplitters, wedges) is 27\,mm and therefore material dispersion temporally stretches the UV pulses. 
The setup is designed in a way that dispersion in both arms is balanced at $\tau =0$\,fs.
A prism-compressor is used to recompress the UV pulses close to their transform-limited duration of $\approx 50$\,fs, which is larger than the duration of the NIR pulses due to phase-matching bandwidth limitations in the THG setup. 
%Note that the phase matching bandwidth of the crystals in the THG setup was not large enough to triple the entire NIR spectrum.

The compressed UV pulses are focused into a gas cell using an $f=200$\,mm CaF\textsubscript{2} lens mounted in vacuum. 
Due to the onset of filamentation in the bulk material of the optics the energy per pump pulse is limited to $\approx 50$\,\textmu J. 
The gas cell used to generate the XUV light has a length of 6\,mm and the focus is placed right at the entrance of the cell.
The third harmonic ($\approx$14\,eV) of the 266\,nm light is generated either in argon or krypton.
% and separated from the majority of the UV light using an aperture. 
The intensity of the XUV pulses is monitored with a microchannel plate (MCP) whose boxcar-averaged output is fed into the lock-in amplifier.
Note that the XUV pulses are not separated from the driving UV pulses optically. Only an aperture in the XUV beamline is used to block some of the UV pump light by exploiting the smaller divergence of the XUV light. Instead, the phase-modulation scheme described in the next section is used to filter the XUV contribution from the total MCP signal electronically.
However, care was taken that the MCP was not saturated by the total light intensity.

\subsection{Phase modulation scheme}
The signals are detected using a phase modulation technique in combination with lock-in detection, which is described in detail in Refs.\,\cite{tekavec_wave_2006, bruder_phase-modulated_2017}. 
We outline the principle here only briefly. 
A detailed calculation of the signals can be found in the Supplementary Information. 
The AOMs are driven at distinct radio frequencies in phase-locked mode ($\Omega_1 = 2\pi \times 160$\,MHz, $\Omega_2 = \Omega_1 +2\pi \times 110$\,Hz).
Bragg-diffraction in the AOM crystal shifts the optical frequency $\omega$ by the AOM frequency: $\omega \rightarrow \omega + \Omega_i$. 
%Hence, photons traveling through different interferometer arms exhibit a relative frequency beating of $\Omega_{21}= \Omega_2- \Omega_1$. 
Hence, the interference of photons from both interferometer arms exhibits a low-frequency beating at a frequency of $\Omega_{21}= \Omega_2 - \Omega_1 = 2\pi \times 110\,$Hz (i.e. linear shot-to-shot phase cycling). 

To track the relative phase of the two interferometer arms a second continuous wave (CW) UV laser beam (Crylas FQCW 266) co-propagates with UV pulses through the interferometer.
We record the interference after the interferometer with a photodiode and the intensity beating of the CW light serves as a reference signal:
\begin{equation} \label{Sref}
    S_\mathrm{ref} \propto 1 + \cos(\omega_0 \tau + \Omega_{21} t)\,.
\end{equation}
For the UV fs pulses traveling through the interferometer, the beating is sampled by the repetition rate of the fs laser. 
Upon third-harmonic generation by the UV pulses in the gas cell, the frequency of the fs pulses is up-converted, leading to an up-conversion of the beat frequency between the XUV pulses to $3\Omega_{21}$, respectively. 
The modulation frequency is chosen such, that the Nyquist sampling criterion is fullfilled: $0.5\times f_\mathrm{rep}=500\,\mathrm{Hz}\gg 3\Omega_{21}/2\pi$, but also the frequency is high enough to efficiently suppress $1/f$-noise\,\cite{bruder_phase-synchronous_2018}.

We record the interfereometric CC of these modulated XUV pulse pairs: 
%Thus the electric field of each UV pulse after passing trough the $i$\textsuperscript{th} ($i = 1,2$) AOM can be written as
%\begin{eqnarray}
%E_i(t) &=& \text{Re}\, A(t) \exp{[i(\omega +\Omega_i)t]}.
%\end{eqnarray}
%Here $A$ denotes the time-dependent envelope of the electric field and $\omega$ is the carrier frequency.
%The phase difference between the two electric fields separated by a delay $\tau$ is
%$\phi_{21} = -\omega\tau + (\Omega_2 - \Omega_1 ) t $, which follows immediately because $\Omega_i \tau \ll \Omega_i t$.
%Hence the relative phase of the UV pulses is modulated at a frequency of $\Omega_{21} := \Omega_2 -\Omega_1$ in the laboratory time frame.
%This modulation is sampled at the repetition rate of the laser system. 

%If the XUV generation is a coherent process, the $\Omega_{21}$-modulation of the UV pulses transfers to a $3\Omega_{21}$-modulation of the XUV pulses.
%In the experiment the linear interferometric CC of the XUV pulse pair, generated by the phase-modulated UV pulse pair is recorded.
%A detailed derivation of the modulated CC signal and its demodulation by lock-in detection is presented in the Supplementary Information. 
%Here only a brief summary of the results is presented.
%Let $A_i(\omega)$ and $\phi_i(\omega)$ be the spectral amplitude and phase of the i\textsuperscript{th} XUV pulse.
%Then the linear interferometric CC of the XUV pulses is given by 
\begin{equation}
S(t,\tau) = W_1 + W_2 + 2 \int \mathrm{d}\omega \, A_1(\omega) A_2(\omega) \cos\left[\omega\tau + \phi_{21}(\omega) + 3\Omega_{21} t\right]\, . \label{eq:Sttau}
\end{equation}
Here, $W_i$ denotes the pulse energy and $A_i$ the spectral amplitude of the XUV pulses ($i=1,2$). 
$\phi_{21}(\omega) = \phi_2(\omega) - \phi_1(\omega)$ denotes the spectral phase difference between both pulses. %The signal oscillates at the XUV carrier frequency with respect to $\tau$ and is modulated at a frequency of $3\Omega_{21}$ in the laboratory time frame.
%Furthermore, $\phi_{21}(\omega) = \phi_2(\omega) - \phi_1(\omega)$.

This signal is demodulated with a lock-in amplifier (Zurich Instruments MFLI), referenced to the continuous reference signal $S_\mathrm{ref}$ of equation\,\ref{Sref}. 
Note that the modulation frequencies are different in the CC signal (equation\,\ref{eq:Sttau}): $3\Omega_{21}$, and reference signal (equation\,\ref{Sref}): $\Omega_{21}$. 
We therefore use third harmonic lock-in detection, which extracts the amplitude and the phase of the CC signal oscillating at the frequency $3\Omega_{21}$ and thus yields the complex demodulated CC signal:
\begin{eqnarray}
\overline{S}(\tau) = \int \mathrm{d}\omega \, A_1(\omega) A_2(\omega) e^{i [(\omega - 3\omega_0)\tau +\phi_{21}(\omega)]}\, .  \label{eq:cc}
\end{eqnarray}

This detection scheme has several advantages. The retrieved signal $\overline{S}$ oscillates at a reduced frequency of $\overline{\omega} = \omega - 3\omega_0$ with respect to the pulse delay $\tau$ (rotating frame detection)\,\cite{tan_theory_2008}. 
Accordingly, signal frequencies are reduced in our experiment by a factor of $\omega/\overline{\omega} \leq 120$, which permits sampling with much larger delay increments and thus reduces data acquisition times.  
%This decreased the acquisition time for a delay-scan by the same amount due to the bigger permissible delay increments for sampling.
Furthermore, phase noise from the interferometer appears correlated in the reference and fs laser signals. 
In the lock-in detection, the phase difference between the third-harmonic reference and the CC signal is computed upon which correlated phase noise cancels out to a large extent.
This has the effect of efficient passive phase stabilization of the interferometer. 
In addition, the lock-in amplifier acts as a very steep bandpass filter combined with high amplification, extracting only signals modulated at a frequency of $3\Omega_{21}$. 
This efficiently suppresses signals e.g. from fundamental light or non-interference signals, providing a background-free output of XUV interference signals. 
%Hence the majority of the jitter arising from the interferometer is removed in the lock-in process.
%A Fourier transform of equation\,\ref{eq:Sttau} yields the spectrum of the XUV pulses, shifted by $3\omega_0$.
%More details on the phase-modulation technique, its combination with harmonic generation and the lock-in detection scheme can be found in Refs.\,\cite{tekavec_wave_2006,wituschek_tracking_2020,bruder_phase-modulated_2017}.
%
\begin{figure}[ht]
\centering\includegraphics[width=0.8\linewidth]{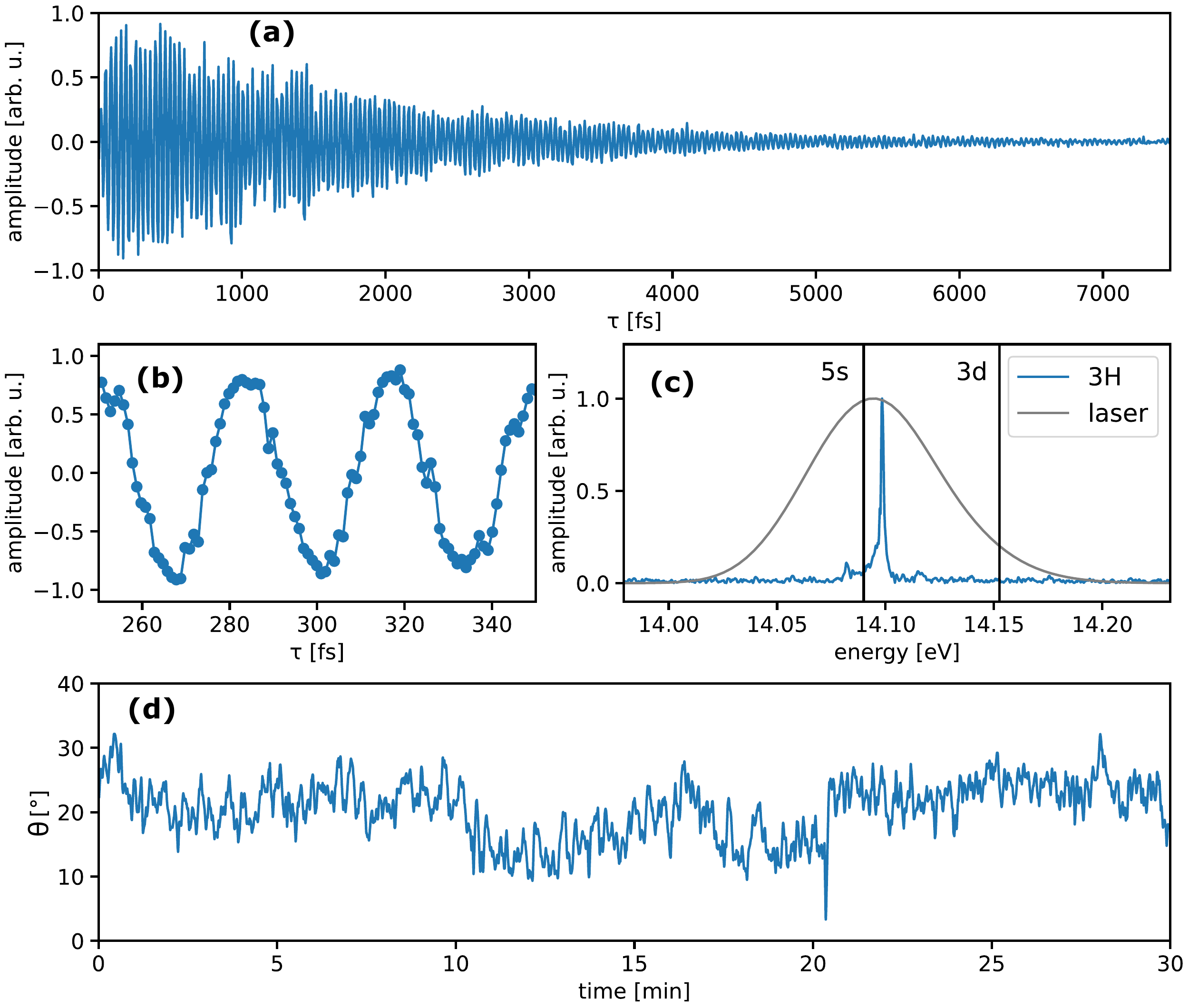}
\caption{XUV generation in argon. (a) Fringe-resolved interferometric CC of the XUV pulses. The curve represents $\text{Re}\lbrace \overline{S}(\tau)\rbrace$. (b) Fine scan of the interferogram with stepsize of 1\,fs. (c) XUV spectrum obtained from Fourier transformation. Black vertical lines indicate the $3p^5(^2P_{3/2})5s~ ^2[3/2]$ and $3p^5(^2P_{3/2})3d~ ^2[3/2]$ resonances respectively. The grey line shows the XUV spectrum which is anticipated by scaling of the UV spectrum. It is much broader than the actually generated spectrum. (d) Phase-stability of the demodulated signal over time.
 }
\label{fig:argon}
\end{figure}

\section{Results}\label{sec:results}
\subsection{Phase modulation of XUV pulses}\label{sec:results_below}
Third harmonic XUV pulses are generated in argon below the ionization threshold at a pressure of 89\,mbar in the gas cell. 
The UV pulse energy is 12\,\textmu J per pulse and the spectrum is centered at 264.8\,nm. 
Note that under these conditions no higher-order harmonics ($>3$) are observed.
Linear interferometric CCs of the resulting XUV pulses, are recorded by scanning the delay between the two UV pump pulses (see figure\,\ref{fig:setup}). 
Consequently, the Fourier transform of the CC trace yields the XUV spectrum.

Figure\,\ref{fig:argon}\,(a) shows the real part of the signal $\text{Re}\lbrace \overline{S}(\tau)\rbrace$ as obtained by third-harmonic lock-in detection (only positive delays are shown).
The data shows clear interference fringes over a range of $\geq$7\,ps.
Due to the rotating frame sampling the oscillation period is $\overline{T}= 2 \pi /\overline{\omega} \approx 35$\,fs instead of $T=2 \pi /\omega \approx 293$\,as that would be expected for $\omega = 14$\,eV$/\hbar$.
This allowed us to record the data shown in figure\,\ref{fig:argon}\,(a) in only 17\,min using a step size of 7\,fs for the delay scan.
Averaging over 660 laser shots was performed at each delay step using a lock-in filter bandwidth of $f_\text{-3dB}=0.4$\,Hz.
In figure\,\ref{fig:argon}\,(b) a fine-scan from 250-350\,fs with a step size of 1\,fs is shown. The data points follow a clean oscillation.
\begin{figure}
\centering\includegraphics[width=0.8\linewidth]{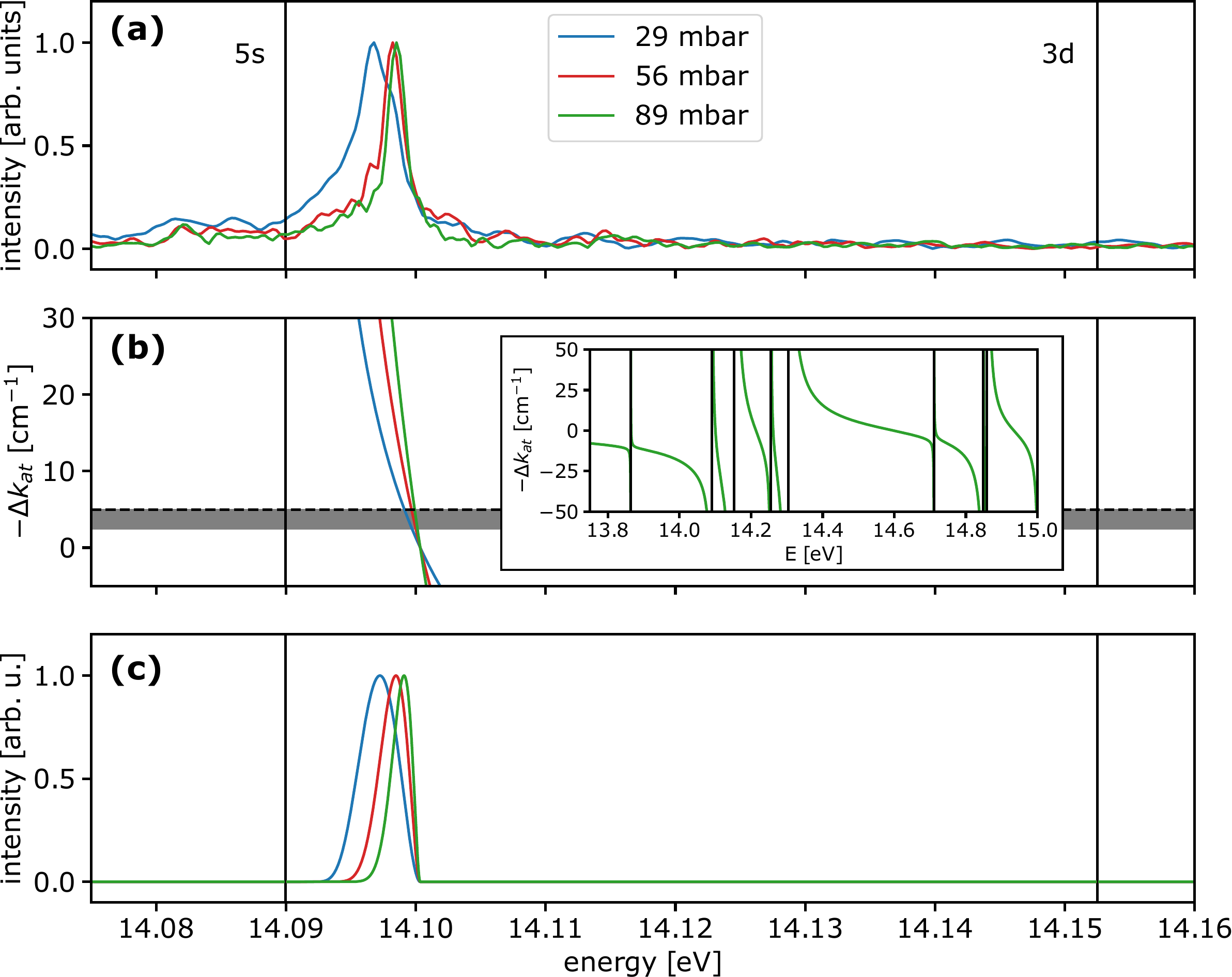}
\caption{(a) XUV spectra generated in argon for different pressures in the gas cell (all normalized to unit amplitude). (b) Calculated phase-mismatch $-\Delta k_{\text{at}}$ for different pressures in the gas cell. The dashed horizontal line indicates $\Delta k_{\text{g}}(0)$, the grey area corresponds to the Rayleigh range. The vertical black lines indicate the position of the nearby resonances. The inset shows $-\Delta k_{\text{at}}$ calculated for a pressure of 89\,mbar for a broader spectral range in argon.
(c) Calculated spectrum of the generated XUV light for different pressures in the gas cell.
}
\label{fig:mismatch}
\end{figure}

The effective phase stability of our experiment is determined using the procedure outlined in Ref.\,\cite{wituschek_stable_2019}.
Here, the phase $\theta = \arg\lbrace \overline{S}(\tau)\rbrace$, obtained by the lock-in amplifier is recorded at a fixed delay of $\tau = 500$\,fs for 30\,min, using a lock-in filter bandwidth of  $f_\text{-3dB}=0.1$\,Hz.
The result is shown in figure\,\ref{fig:argon}\,(d).
The standard deviation from the mean value is only $\delta\theta =4.5^\circ$ over 30\,min, which is a remarkable result, considering the high photon energy of $\approx$\,14\,eV.
Assuming that the phase drift $\delta\theta$ is caused by a delay drift, we can calculate the corresponding delay drift $\delta\tau = \frac{\delta \theta}{\omega -3\omega_0}\approx 457$\,as, for $\omega = 14.09\,\mathrm{eV}/\hbar$. This is equivalent to a path-length drift of 137\,nm.
However, the path-length drifts of the interferometer on the timescale of 30\,min are on the order of 6.3\,nm (see Supplementary Information).
Therefore, the effective phase stability $\delta\theta$ is limited by amplitude fluctuations in the signal, rather than by the stability of the interferometer.

By taking the Fourier transform of the time-domain interferogram the spectrum of the XUV pulses is obtained with a spectral resolution of 0.65\,meV (FWHM) [figure\,\ref{fig:argon}\,(c)].
One can clearly see a well-resolved, narrow spectral peak (FWHM = 1.7\,meV) appearing between the $3p^5(^2P_{3/2})5s~ ^2[3/2]$ and $3p^5(^2P_{3/2})3d~ ^2[3/2]$ resonances. 
The frequency axis is calibrated by the known reference laser frequency $\omega_0 = 2\pi \times 1.12653(5) \times 10^{15}$\,rad\,s\textsuperscript{-1} and the spectrum is corrected for the $3\omega_0$-shift due to rotating frame sampling.
The obtained XUV spectrum is much narrower than what would be expected from a non-resonant process, i.e. by scaling the UV spectrum $I(\omega)$ with the harmonic order according to $I(\omega)^3$ [see grey curve in figure \,\ref{fig:argon}\,(c)]. $I(\omega)$ is measured with a fiber spectrometer.

The narrow spectrum can be explained by enhanced phase matching occurring on the high-energy side of an atomic resonance\,\cite{mahon_third-harmonic_1979}. 
In a simple model, the phase mismatch between the generated harmonic and the laser-induced polarization is given by $\Delta k = \Delta k_{\text{at}} + \Delta k_\text{g}$, where $\Delta k_{\text{at}}$ is the phase mismatch of the neutral atomic medium and $\Delta k_\text{g}$ is the Gouy-phase shift:

\begin{eqnarray}\label{eq:1}
\Delta k_{\text{at}}(E)	&=& \frac{2\pi}{h c } E (n_\text{XUV} - n_\text{UV}) \\
\Delta k_{\text{g}}	(z)	&=& \frac{q-1}{2\pi} \frac{\mathrm{d}}{\mathrm{d} z} \arctan(z/z_R)
\end{eqnarray}

Here $n_\text{XUV}$ and $n_\text{UV}$ are the refractive indexes of the neutral generation medium in the XUV and the UV, respectively, $E$ is the photon energy, $h$ is the Planck constant, $c$ is the speed of light, $q$ is the order of the harmonic, $z$ the position along the propagation axis of the laser relative to the focus position and $z_R$ is the Rayleigh-range in the focus.
The pressure-dependent refractive index can be calculated with Sellmeier's equation\,\cite{mahon_third-harmonic_1979}, using the state energies and transition dipole moments reported in Refs.\,\cite{kramida_NIST_2019, chan_absolute_1992}.
If $\{E_j\}$ are the resonances of the medium, then for $E>E_j$, $\Delta k_{\text{at}}$ can be negative and therefore compensates the phase mismatch induced by the Gouy-phase shift.
The spectrum of the generated XUV light is determined by\,\cite{mahon_third-harmonic_1979}
%\begin{eqnarray}\label{eq:P3}
%P_\text{XUV} &\propto& E^4 \chi^{(3)}(E)^2 P_\text{UV}^3  b^2 \Delta k^2 \exp(b\Delta k), ~\text{for}~ \Delta k <0\\
%    &=& 0, ~\text{for}~ \Delta k \geq 0,
%\end{eqnarray}
\begin{eqnarray}\label{eq:P3}
P_\text{XUV}(E) \begin{cases} \propto E^4 \chi_{(3)}^2(E) P_\text{UV}^3  b^2 \Delta k^2 \exp(b\Delta k)\; &\text{for}\; \Delta k <0\\
    = 0\; &\text{for}\; \Delta k \geq 0,
\end{cases}
\end{eqnarray}
where $\chi_{(3)}$ is the third-order nonlinear susceptibility, $P_\text{UV}$ is the power of the fundamental and 
$b = \frac{2 \pi w_0^2 }{\lambda} \approx 0.16$\,cm is the confocal parameter of the UV laser, determined by the waist size $2w_0$ and the UV laser carrier-wavelength $\lambda$.

In figure\,\ref{fig:mismatch}\,(a) we show spectra of the generated XUV pulses acquired for different pressures in the gas cell.
We observe a blue-shift and a narrowing of the spectrum with increasing pressure. 
This is consistent with calculations of the atomic phase-mismatch $\Delta k$ and the resulting XUV spectrum, which are shown in figure\,\ref{fig:mismatch}\,(b) and (c).
For the calculations, we assume a flat $\chi_{(3)}$ between the resonances, where the generated XUV light appears.
Furthermore, the Rayleigh range is used as a parameter in the calculations. 
It coincides with the Rayleigh range which can be determined from the experimental parameters.
From the calculations one can see that phase-matching shifts to higher energies when the pressure increases.
In addition, the spectral width of the phase-matching region depends on the gradient of $-\Delta k_{\text{at}}$ within the Rayleigh range.
The gradient increases with increasing pressure, hence the spectral width of the XUV decreases.
The experimental data and the calculations agree very well, which is remarkable considering the high spectral resolution, exceeding other studies on resonance-enhanced phase matching by a factor of $> 60$\,\cite{chini_coherent_2014}.

\begin{figure}
\centering\includegraphics[width=0.8\linewidth]{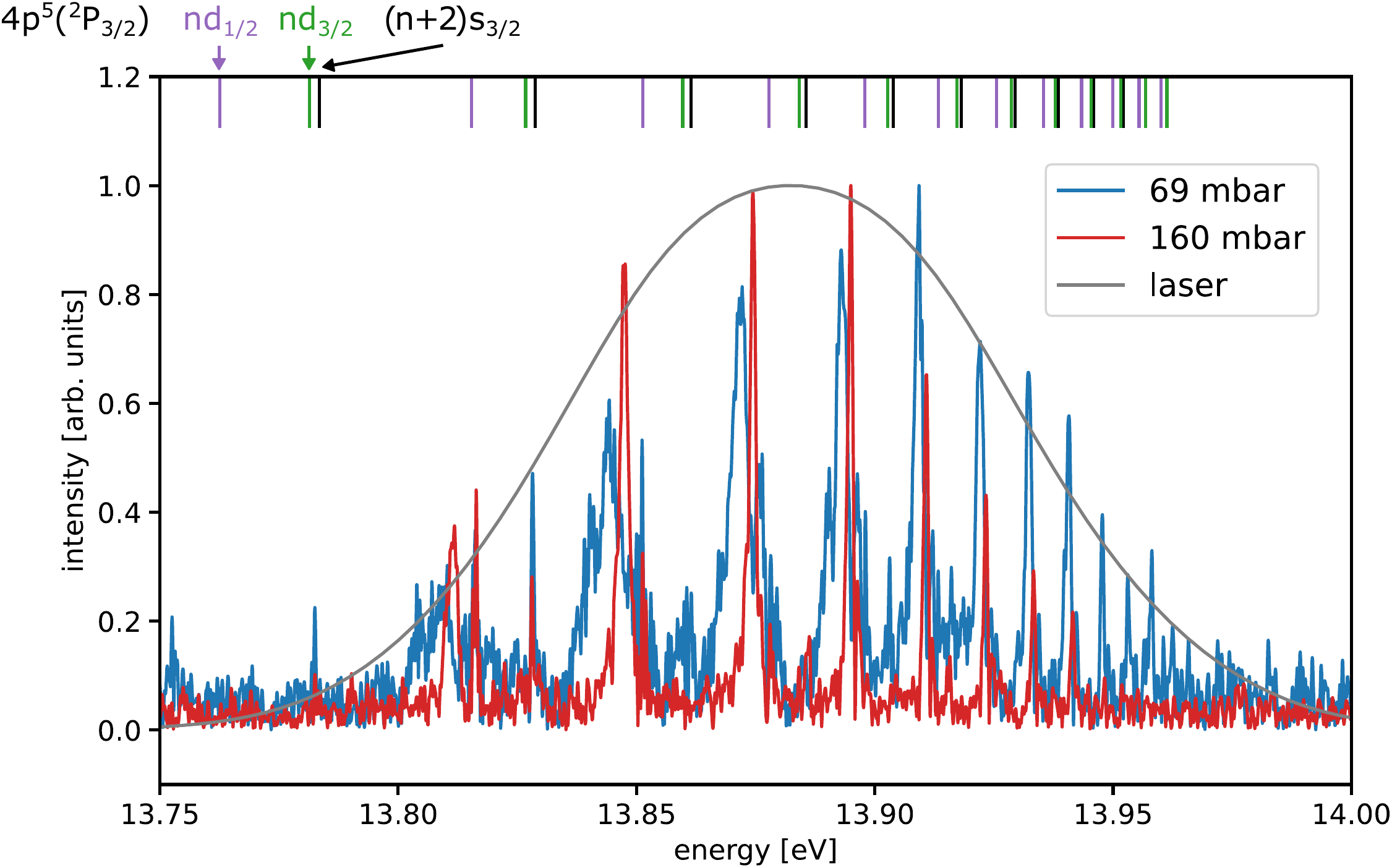}
\caption{Normalized spectra of the XUV light generated in krypton for two different pressures. The grey line shows the XUV spectrum which is anticipated by scaling of the UV spectrum. The positions of the Rydberg series that participate in the XUV generation are indicated at the top of the figure. These Rydberg series converge to the $4p^5(^2P_{3/2})$ threshold at 14\,eV.
}
\label{fig:krypton}
\end{figure}

The measurement was repeated using krypton as the generation medium.
The UV pulse energy was set to 20\,\textmu J per pulse and the spectrum was centered around 268.1\,nm.
In contrast to the argon experiment not only two resonances but several resonances of a Rydberg-series, converging to the $4p^5(^2P_{3/2})$ limit, occurr within the bandwidth of the UV laser.
Consequently, phase-matching is achieved at several frequencies between the neighboring resonances, resulting in a complex spectral structure\,\cite{drescher_xuv_2020} (see figure\,\ref{fig:krypton}).
Similar to the argon experiment a blue-shift of the peak positions and a spectral narrowing is observed when increasing the pressure in the gas cell. 
By scanning the delay from 0\,ps to 9\,ps a resolution of 0.5\,meV FWHM is obtained by the Fourier-transform.
%Note that the side-peaks and narrow spectral spikes in the spectrum of figure\,\ref{fig:krypton} are not artifacts or noise from the measurement but instead a result of processes occurring during the XUV generation. 

\subsection{Double pump-pulse effects}\label{sec:overlap}
So far, we analyzed the data for pulse delays larger than the pulse duration of the UV driving pulses. This was sufficient to analyze the spectral shape of the generated narrow-band XUV pulses. We now turn our focus on effects occurring when the two driving UV pulses overlap temporally and spatially in the harmonic generation medium. 

The perspective of generating phase-locked pulse pairs with collinear pulse sequences is appealing due to the simplicity of such setups. 
However, a shortcoming is the issue that the perturbation of the harmonic generation medium by the leading pulse (optical excitation and ionization of the medium), may change the amplitude and phase of the XUV radiation generated by the trailing pulse. 
This issue plays also an important role in HHG at high repetition rates $>10$\,MHz\,\cite{porat_phase-matched_2018}. 
Previous studies investigated this effect by analyzing the contrast of interference fringes detected with XUV spectrometers\,\cite{bellini_phase-locked_2001, salieres_frequency-domain_1999}. 
Here, we study this effect in the time domain for the case of phase-matching enhanced harmonic generation near atomic resonances. As shown in the Supplementary Information, our CC measurements provide direct access to the relative phase between the XUV pulses without the need of theoretical assumptions. 
Yet, information about the individual amplitude of the pulses cannot be obtained due to the intrinsic symmetry of CC measurements (cf. Supplementary Information). We therefore concentrate on the relative phase shift induced by the perturbation of the harmonic generation medium. 

%The previous sections provide analysis of interferograms recorded at interpulse delays, which exceed the duration of the driving UV pulses. 
%When the two UV pulses overlap in the medium, harmonic generation is modified by interference between them.
%As previously demonstrated, the harmonic generation from two driving pulses can be assumed to be independent from each other as long as the leading pulse induces negligible change in the properties of the generation medium\,\cite{bellini_phase-locked_2001}.
%Such a change can be ionization or excitation of the medium. 
%In the case of a significant modification of the generation medium due to excitation or ionization, the trailing pulse may generate harmonics with a different phase and amplitude compared to the harmonics created by the leading pulse. Here, we record interferometric CCs of the generated XUV pulses for different pulse energies to specifically study the change in phase shift induced between the pulses. 

\begin{figure}
\centering\includegraphics[width=0.75\linewidth]{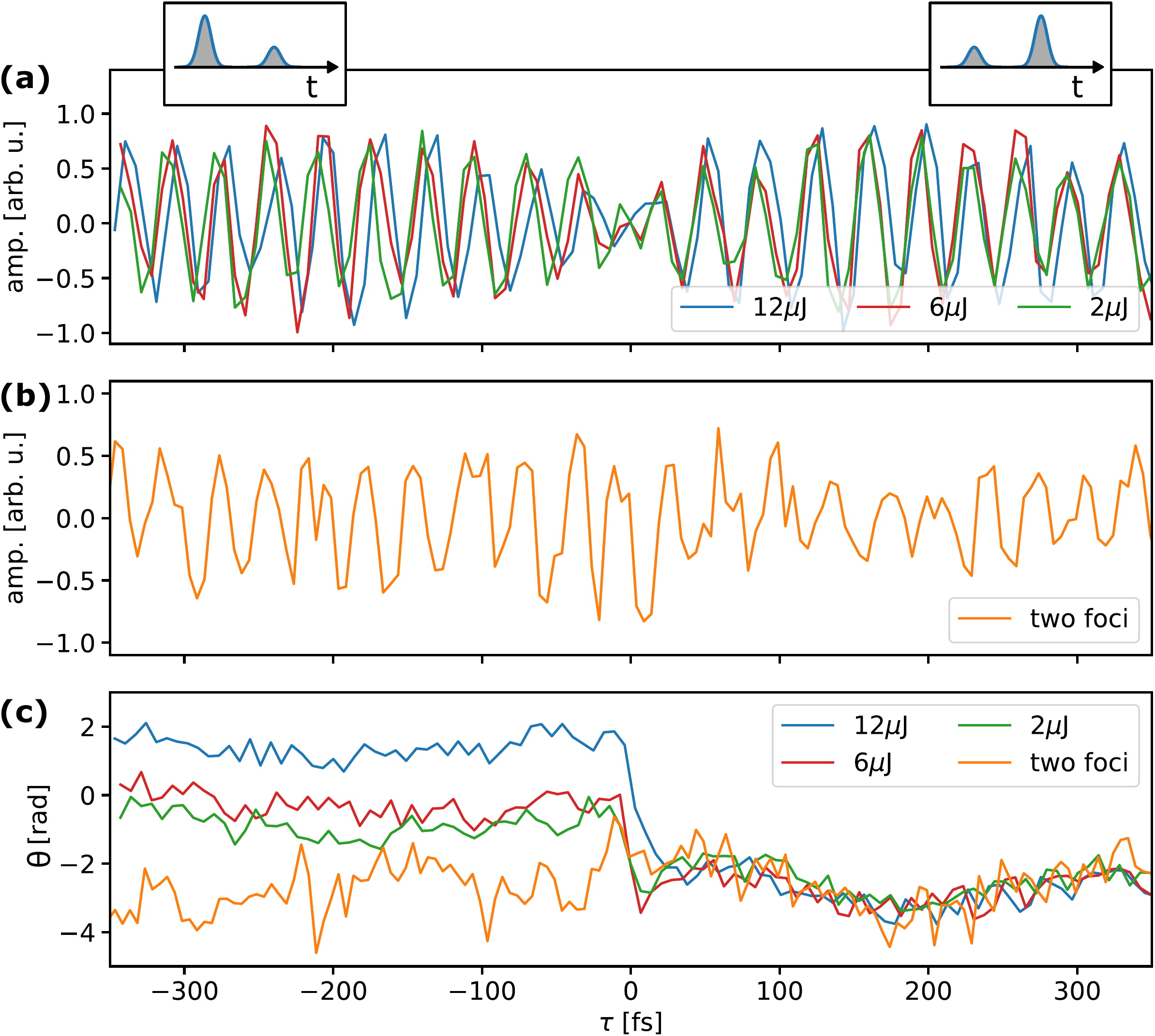}
\caption{ Comparison of interferometric CCs using variable pump energy in one of the two UV pulses. (a) Traces for three different values of $E_\text{var}$. For negative delays the variable energy pulse comes first, for positive delays the fixed energy pulse comes first (b) Both pump pulses exhibit equal pulse energy and use spatially separated volumes for harmonic generation. Note that the curves in (a) and (b) represent $\text{Re}\lbrace \overline{S}(\tau)\rbrace$.
(c) Phase $\theta = \arg \lbrace \overline{S}(\tau) \rbrace$ of the oscillations of the interferograms shown in (a) and (b). For the blue, red and green curve the phase has been unwrapped from the right side (positive delays) and from the left side (negative delays) in order to obtain the correct phase jump at zero delay. In order to make the phase differences more visible, the linear phase evolution was subtracted from the unwrapped phase. Clear phase shifts are observed between positive and negative delays. For the two-foci data (orange) no such phase shift is observed.}
\label{fig:power}
\end{figure}

To this end, we systematically attenuate the pulse energy of one of the two UV pump-pulses ($E_\text{var} =$ 2\,\textmu -12\,\textmu J) while keeping the energy of the other pulse fixed ($E_\text{fix} =$ 12\,\textmu J) and record interferometric CCs (figure\,\ref{fig:power}). 
The power is adjusted by reducing the RF driving power of one of the AOMs.
All other XUV generation parameters are the same as for the argon study at a gas cell pressure of 89\,mbar.
%The results are shown in figure\,\ref{fig:power}.
%The pulse energy in one arm ($E_\text{var} =$ 2\,\textmu -12\,\textmu J) was adjusted and kept fixed in the other arm ($E_\text{fix} =$ 12\,\textmu J). 
%Adjustment was done by reducing the RF driving power of one AOM.
At positive delays, the pulse with fixed energy interacts with the harmonic generation medium first.
%comes first, all CC interferograms are in phase [see figure\,\ref{fig:power}\,(c)]. 
Hence, for positive delays, the amount of perturbation induced by the leading pulse is the same for all measurements and a constant phase shift between the XUV pulses is expected. This is in agreement with our data, where all CC interferograms are in phase for positive delays [figure\,\ref{fig:power}(a, c)]. 
Note, that in these considerations, phase shifts introduced by the trailing pulse itself when changing its intensity can be neglected.
In contrast, for negative delays, where the variable-energy pulse comes first, the perturbation of the generation medium by the leading pulse changes between the measurements.
Therefore the second pulse experiences different conditions, leading to systematic phase-shifts depending on $E_\text{var}$, as observed in our data for negative delays [figure\,\ref{fig:power}\,(a, c)]
The phase shift is expected to vanish for large negative delays on timescales given by the decay of the excitation/ionization in the generation medium.

The influence of the leading pulse on the harmonic generation from the trailing pulse can be omitted when the harmonic generation takes place in two spatially separated volumes of the generation medium, as previously demonstrated\,\cite{ jansen_spatially_2016,bellini_temporal_1998,descamps_extreme_2000, kovacev_extreme_2005}. 
However, this leads to a reduced spatial overlap of both beams in the far field, making it more challenging to achieve good signal-to-noise performance. 
Here, the phase modulation scheme is of advantage in maintaining the signal interference contrast as it suppresses non-interfering signal contributions.  
In figure\,\ref{fig:power}\,(b) results are shown that were obtained under conditions where the foci are separated by $\approx$\,100\,\textmu m and both pulses have equal energy. 
The separation of the two foci is realized by slightly misaligning the interferometer. 
In this case no phase shifts occur in the interferogram [see figure\,\ref{fig:power}\,(c)], when going from negative to positive delays.
The lower signal quality in this data originates mainly from the reduced signal amplitude due to the reduced spatial overlap of the beams. 
%the lower interference contrast in the non-collinear geometry. Yet, the signal quality is still decent which is a consequence of the efficient background suppression in the phase modulation scheme suppressing non-interfering signal contributions. 

Note that the different temporal position of the phase jump in figure\,\ref{fig:power}\,(c) around zero delay stems from different discretization of the delay axis in the measurements, and that the magnitude of the phase jump is measured modulo $2\pi$.
%However, the fact that the amplitude of the oscillations is equal for positive and negative time delays suggests that the total XUV pulse energy depends only the product $E_\text{fix} E_\text{var}$.
Furthermore, $\theta = \arg \lbrace \overline{S}(\tau) \rbrace$ has an offset-calibration uncertainty due to the different phase-transfer functions of the electronics (for example photo-diodes, amplifiers) used in the signal and reference pathways (remember that the lock-in amplifier measures the phase-difference between the signal and reference inputs).
This explains why the curve in figure\,\ref{fig:power}\,(b) is not symmetric with respect to $\tau=0$.
We did not perform a phase calibration in this study, however with a straightforward calibration also the absolute phase scale could be retrieved\,\cite{bruder_delocalized_2019}.

\section{Discussion and conclusion}
We successfully generated phase-locked XUV pulse-pairs and gained independent control over the delay and relative phase of the XUV pulses through manipulation of the pump pulses.
This novel level of control was reached by implementing a phase-cycling scheme for the XUV pulses, which allowed for pathway-selective detection of individual signal contributions and rotating frame detection, efficiently reducing data acquisition times. 
Here, the pathway-selective detection selects a single harmonic and isolates the interference signals from non-modulated background signals. 
%Here, pathway-selectivity in the detection was achieved by means of blocking any signals without a distinct modulation with the lock-in amplifier.
This is especially beneficial for photoion/-electron spectroscopy in the XUV-regime, where many background events due to one-photon XUV ionization are present. 
%These events exhibit no modulation outside the temporal overlap region of the XUV pulses and thus will be blocked by the lock-in amplifier.

In addition, due to the phase modulation technique the majority of the phase jitter originating from the interferometric pump-pulse setup is efficiently removed in the lock-in demodulation process, yielding a phase-stability of 4.5$^\circ$ over 30\,min.
Note that this stabilization method is passive and therefore requires no active stabilization of the interferometer optics.

Due to the high resolution of the Fourier-transform approach (0.5\,meV FWHM in the krypton study) we were able to measure gas-pressure dependent spectral widths and shifts of narrow-band XUV light generated in argon and krypton.
This is a much higher resolution than achieved in typical XUV Fourier transform spectroscopy studies. 
Our data is in good agreement with a theoretical model, showing that the narrow-bandwidth generation is a consequence of enhanced phase-matching at the high-energy side of atomic resonances in the generation medium.
In the argon study a single peak emerged in the spectrum that is much narrower than the bandwidth that would be anticipated from the pump laser spectrum.
In the krypton study phase matching enhanced the generation of several narrow bandwidth features that appear between the closely spaced resonances of a Rydberg series.
We note that in contrast to this study, intense narrow bandwidth XUV pulses can also be generated by resonance-enhanced harmonic generation exploiting Stark-shifted multi photon resonances\,\cite{ackermann_resonantly-enhanced_2012}.

The high spectral resolution of our work, resolving peak positions, widths and shapes may foster and challenge detailed theoretical studies including quantum models.
Note that in the Fourier-transform approach, the resolution limit is given by the inter-pulse delay range.
With XUV Ramsey-comb spectroscopy even higher resolution in the MHz range may be achieved\,\cite{kandula_extreme_2010}.

By looking at pump-intensity dependent phase-shifts in the CC measurements it was found that in our experimental conditions the XUV generation from the individual UV pulses was not independent and leads to phase-shifts in the XUV pulse generated by the trailing UV pulse.
These intensity dependent phase-shifts were directly measured by our all-optical approach. 
Furthermore, we could confirm that perturbations in the harmonic generation medium by leading pulses can be omitted using spatially well-separated foci for both driving pulses\,\cite{bellini_phase-locked_2001}. 
%In this view, we also generated XUV pulse pairs using spatially separated foci.
%For this case, we obtained still a decent signal quality while the phase-shift disappeared, indicating well-separated foci. 

Our results significantly extend and improve existing XUV time-domain spectroscopy schemes\,\cite{cavalieri_ramsey-type_2002, mang_spatially_2018} by uniquely combining pathway-selective detection with high phase-stability and high spectral resolution. 
As such, this work may pave the way for many applications of (non)linear coherent time-domain spectroscopy in the XUV spectral range.
A straightforward extension to higher harmonics will provide access to higher photon energies and shorter pulse durations, and in combination with additional probe pulses, dynamics may be probed with high spectro-temporal resolution. 

\section*{Funding}
Funding by the Bundesministerium für Bildung und Forschung (BMBF) \textit{STAR} (05K19VF3), by the European Research Council (ERC) Advanced Grant \textit{COCONIS} (694965), and by the Deutsche Forschungsgemeinschaft (DFG) GRK 2079 and STI 125/24-1 is acknowledged. 

\section*{Acknowledgments}
We gratefully acknowledge the support of Ahmet Akin Uenal and Evgenii Ikonnikov in preparing the laser system and setting up the XUV-beamline.

%%%%%%%%%% Bibliography:
\section*{References}

\bibliography{literature.bib}

\end{document}

%Title of paper
\title{Supplementary Material}
% \author{Andreas Wituschek}
% \affiliation{Institute of Physics, University of Freiburg, Hermann-Herder-Str. 3, 79104 Freiburg, Germany}
% \author{Oleg Kornilov}
% \author{Tobias Witting}
% \author{Laura Maikowski}
% \affiliation{Max-Born-Institut, Max-Born-Str. 2A, 12489 Berlin, Germany}
% \author{Frank Stienkemeier}
% \affiliation{Institute of Physics, University of Freiburg, Hermann-Herder-Str. 3, 79104 Freiburg, Germany}
% \author{Marc J.J. Vrakking}
% \affiliation{Max-Born-Institut, Max-Born-Str. 2A, 12489 Berlin, Germany}
% \author{Lukas Bruder}
% \email{lukas.bruder@physik.uni-freiburg.de}
% \affiliation{Institute of Physics, University of Freiburg, Hermann-Herder-Str. 3, 79104 Freiburg, Germany}

% \date{November 02, 2020}

\maketitle

\section*{Derivation of the cross-correlation signal}
To clarify the experimental phase-modulation procedure, we derive here exemplary the CC signal. 
We use the following notation:

\begin{itemize}
    \item $\omega$ - optical frequency
	\item $\omega_0$ - optical frequency of the CW reference laser
	\item indexes $i=1,2$ indicate the contributions from the two different interferometer arms
	\item $\Omega_i$ - frequencies of AOM 1 and 2, it is $\Omega_{21} = \Omega_2-\Omega_1$
	\item $R_i(t) = R_i e^{i\omega_o t}$, CW reference field of the interferometer arm $i$, with constant amplitude $R_i$
	\item $\tau$ - pulse delay
	\item $A_i(\omega), \phi_i(\omega)$, $\Delta t$ - spectral amplitudes, phases and the pulse duration of the two XUV pulses
	\item $E_i(t) = \mathcal{F}_{\omega } \left\lbrace A_i(\omega )e^{i\phi_i(\omega )}\right\rbrace$, pulsed XUV fields. %$t'=nT_\mathrm{rep}$ is a quasi-continuous time variable, which increments the carrier-envelope phase of the pulses each laser cycle.
\end{itemize}

In the experiment, the AOMs induce a clean, linear phase shift ($\Delta \phi_\mathrm{AOM}=\Omega_i$) of the transmitted electric fields. This applies for the femtosecond pulses as well as for the CW tracer laser. 
For the monochromatic reference laser we obtain accordingly
\begin{equation}
    R_i(t) = R_i e^{i(\omega_o t + \Omega_i t)}\, .
\end{equation}
Upon harmonic generation of the femtosecond pulses in the gas cell, the modulation is up-shifted to $3\Omega_i t$. Hence, the phase-modulated XUV pulses are described by
\begin{equation}
    E_i(t) = \mathcal{F}_{\omega } \left\lbrace A_i(\omega )e^{i\phi_i(\omega )+3\Omega_i t}\right\rbrace\, .
\end{equation}
First, we calculate the reference signal used for the lock-in detection. 
The reference signal is given by the interference of the two modulated CW beams detected by a photodiode, integrated over the photodiode response time:
\begin{eqnarray}
S_\text{ref} &=& \left< \left| R_{1} e^{i(\omega_0 t + \Omega_1 t)} + R_{2} e^{i(\omega_0 (t+\tau) + \Omega_2 t)} \right|^2 \right>_{det} \nonumber \\ 
&\approx& const + 2 R_{1} R_{2} \cos(\omega_0 \tau + \Omega_{21} t)
\end{eqnarray}
Here, we assumed that the phase modulation $\Omega_{21}t$ is much slower than the detector response time and can be thus regared as a constant phase term over the detector integration time.
Note that within the digital electronics of the lock-in amplifier arbitrary harmonics of this reference signal can be synthesized.

The input signal of the lock-in is given by the interference of the two modulated XUV pulses. The signal is integrated over the time window $T$ of the boxcar averager:
\begin{eqnarray}
S &=& \left< \left| E_{1}(t) + E_2(t+\tau)\right|^2 \right>_{det} \nonumber \\
&=& \int_{-T/2}^{T/2}\mathrm{d}t\, \left| E_{1}(t) + E_2(t+\tau)\right|^2  \nonumber
\end{eqnarray}
Considering $1/\Omega_1\gg \Delta t$, the phase-modulation term can be regarded as a constant term over the pulse duration and hence can be regarded as a constant phase term for the temporal integration over the pulse envelopes. Furthermore, it is $T \gg \tau$, $T \gg \Delta t$ and thus we can set $T \rightarrow \infty$. We can then use Parseval's theorem to calculate the signal:
\begin{eqnarray}
S(t,\tau) &=& \int_{-\infty}^{\infty} \mathrm{d}t \, \left| \mathcal{F}_\omega \left\lbrace A_1(\omega )e^{i(\phi_1(\omega )+ 3\Omega_1 t)} + A_2(\omega )e^{i(\phi_2(\omega )+ 3\Omega_2 t - \omega \tau)}\right\rbrace \right|^2  \nonumber \\
&=& \int_{-\infty}^{\infty} \mathrm{d}\omega\, \left| A_1(\omega )e^{i(\phi_1(\omega )+ 3\Omega_1 t)} + A_2(\omega )e^{i(\phi_2(\omega )+ 3\Omega_2 t - \omega \tau)} \right|^2  \nonumber \\
&=& W_1 + W_2 + \int \mathrm{d}\omega\, 2A_1(\omega)A_2(\omega) \cos(\omega \tau + \phi_{21}(\omega) + 3\Omega_{21}t) 
\end{eqnarray}
In this expression the first two terms correspond to the energies $W_1$ and $W_2$ of the two pulses. $\phi_{21}(\omega) = \phi_2(\omega)-\phi_1(\omega)$ denotes the spectral phase difference of the individual XUV pulses.

Repeating the experiment over many laser cycles leads to a quasi-continuous modulation of the signal at the frequency $3\Omega_{21}$ in the laboratory time frame, which is demodulated using a lock-in amplifier. %For a lock-in time constant $\tau_{LI} \gg T_\mathrm{rep}$, $t'$ can be regarded as a continuous time variable $t' \rightarrow t$. 
The lock-in amplifier measures the Fourier amplitude and phase of $S$ at a given frequency $\Omega$. In the experiment we use third-harmonic lock-in detection with respect to the reference signal $S_\text{ref}$, which is modulated at a frequency $3\Omega_{21}$ and has a phase offset of $3\omega_0 \tau$. 
The latter phase offset is responsible for the `rotating frame' detection. Accordingly, the lock-in signal is
\begin{eqnarray}
\overline{S} (\tau) &=& 
\mathcal{F}_{t} \left\lbrace \int \mathrm{d}\omega\, 2 A_1(\omega) A_2(\omega) \cos\left[(\omega-3\omega_0) \tau + \phi_{21}(\omega)+ 3\Omega_{21} t \right]  \right\rbrace  \nonumber \\
& = & \int  \mathrm{d}\omega\, A_1(\omega) A_2(\omega) e^{i((\omega-3\omega_0) \tau + \phi_{21}(\omega))}  
\end{eqnarray}
%where we considered only the positive frequency spectrum $\Omega > 0$. 
As expected, this signal corresponds to the linear, interferometric cross-correlation signal of the two XUV pulses, detected in the rotating frame of $3\omega_0 \tau$. 
We perform a Fourier transform of the demodulated signal with respect to $\tau$: 
\begin{eqnarray}
\overline{S} (\omega) &=&  \mathcal{F}_\tau\left\lbrace \overline{S}\right\rbrace \nonumber \\
&=& \int \mathrm{d}\tau \int \mathrm{d}\omega' A_1(\omega') A_2(\omega') e^{i\left[(\omega'-3\omega_0- \omega) \tau + \phi_{21}(\omega')\right]} \nonumber \\
&=& \int \mathrm{d}\omega' A_1(\omega' + 3\omega_0) A_2(\omega' + 3\omega_0) e^{i\phi_{21}(\omega' + 3\omega_0)} \underbrace{\int \mathrm{d}\tau \, e^{i(\omega' - \omega)\tau}}_{=\delta(\omega-\omega')} \nonumber \\ 
&=& A_1(\omega + 3\omega_0) A_2(\omega + 3\omega_0) e^{i\phi_{21}(\omega + 3\omega_0)}
\end{eqnarray}
%
%So the experiment measures the overlap between the spectra of the generated XUV pulses.
As expected, for equal amplitude and phase of the XUV pulses ($A_1 = A_2$, $\phi_{21}=0$), the Fourier transform of the cross correlation corresponds to the spectrum of the XUV pulses, however, here down-shifted in frequency by $3\omega_0$ due to rotating frame detection. A shift of the frequency axis by the known reference laser frequency ($+3\omega_0$) recovers the original frequency spectrum. 
Note that for the phase difference $\phi_{21}$ between both XUV pulses can be directly extracted from the signal. 

\section*{Phase stability of the interferometer}
\begin{figure}
\centering\includegraphics[width=0.7\linewidth]{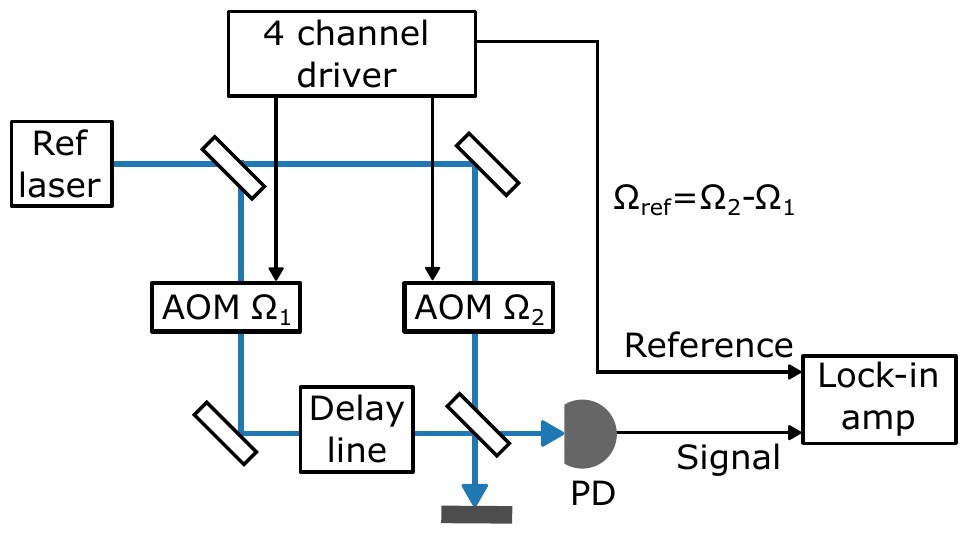}
\caption{Setup for the measurement of the phase stability of the interferometer. Explanations are in the text.
}
\label{fig:setup_phase_stability}
\end{figure}

In order to evaluate the phase stability of the interferometric setup, only the cw reference laser at a wavelength of $\lambda \approx 266$\,nm was coupled into the interferometer (see figure\,\ref{fig:setup_phase_stability}).
At the exit both arms overlap collinearly and their interference is recorded using a photodiode (PD). 
A phase-locked four channel radio-frequency driver provides the two frequencies for the AOMs ($\Omega_1 = 2\pi \times 160$\,MHz and $\Omega_2 = 2\pi \times 160.1$\,MHz), as well as a third reference waveform $\Omega_\text{ref} = \Omega_2 - \Omega_1$.
The AOMs shift the frequency of the reference laser by $\Omega_i$, which leads to a beating in the interference signal at a frequency of $\Omega_2 - \Omega_1$.
This signal is demodulated with a lock-in amplifier using the reference created by the four channel driver. 
The lock-in measures the phase-difference $\theta$ between the interference signal from the PD and the reference waveform, which directly corresponds to the phase difference between the two arms of the interferometer.

\begin{figure}
\centering\includegraphics[width=1\linewidth]{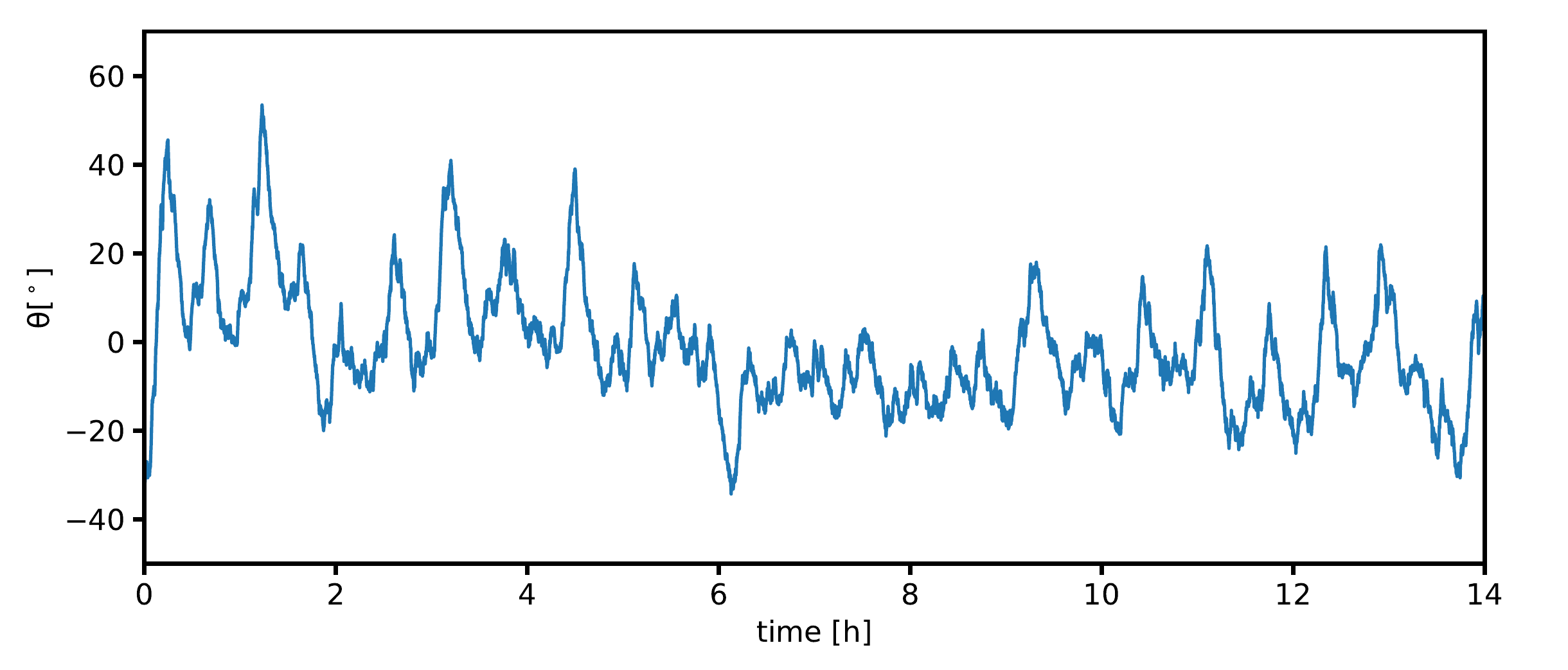}
\caption{Phase stability of the interferometer recorded for 14\,h.}
\label{fig:phase_stability}
\end{figure}

A long term measurement where $\theta$ was recorded for 14\,h is shown in figure\,\ref{fig:phase_stability}. The standard deviation of $\theta$ over 14\,h is $\delta\theta = 13.7^\circ$, while the standard deviation over 30\,min is $\delta\theta_{30} = 8.5^\circ$. 
Assuming that the phase drifts are caused by path-length drifts $\delta L$ in the interferometer, it is $\delta L = \lambda \frac{\delta\theta_{30}}{360^\circ} \approx 6.3$\,nm.

